\documentclass[a4paper,12pt]{article}

\usepackage{multicol}
\usepackage{graphicx}
\usepackage{amsmath}
\usepackage{amsfonts}
\usepackage{amssymb}
\begin{document}
\begin{titlepage}
\begin{flushright}
{\small\begin{tabular}{r}
UWThPh-2001-19\\
April 2001
\end{tabular}}\\
\end{flushright}
\center{
\LARGE{\textbf{The EPR-paradox in massive systems}}\\
\LARGE{\textbf{or about strange particles}}\\
\vspace{1cm}
\large{R.A. Bertlmann, W. Grimus and B.C. Hiesmayr\footnote{Email: hies@thp.univie.ac.at}}\\
\small{\emph{Institute for Theoretical Physics, University of Vienna}}\\
\small{\emph{Boltzmanngasse 5, A-1090 Vienna, Austria}}\\
\vspace{1cm}
\normalsize{\textbf{Abstract}}}\\
\flushleft{\normalsize{We give an introduction to an entangled massive system, specifically the
neutral kaon system,
which has similarities to the entangled two photon system, but, however, also
challenging differences.}}
\vspace{0.5cm}
\center{\includegraphics[width=154.990mm,height=108.813mm]{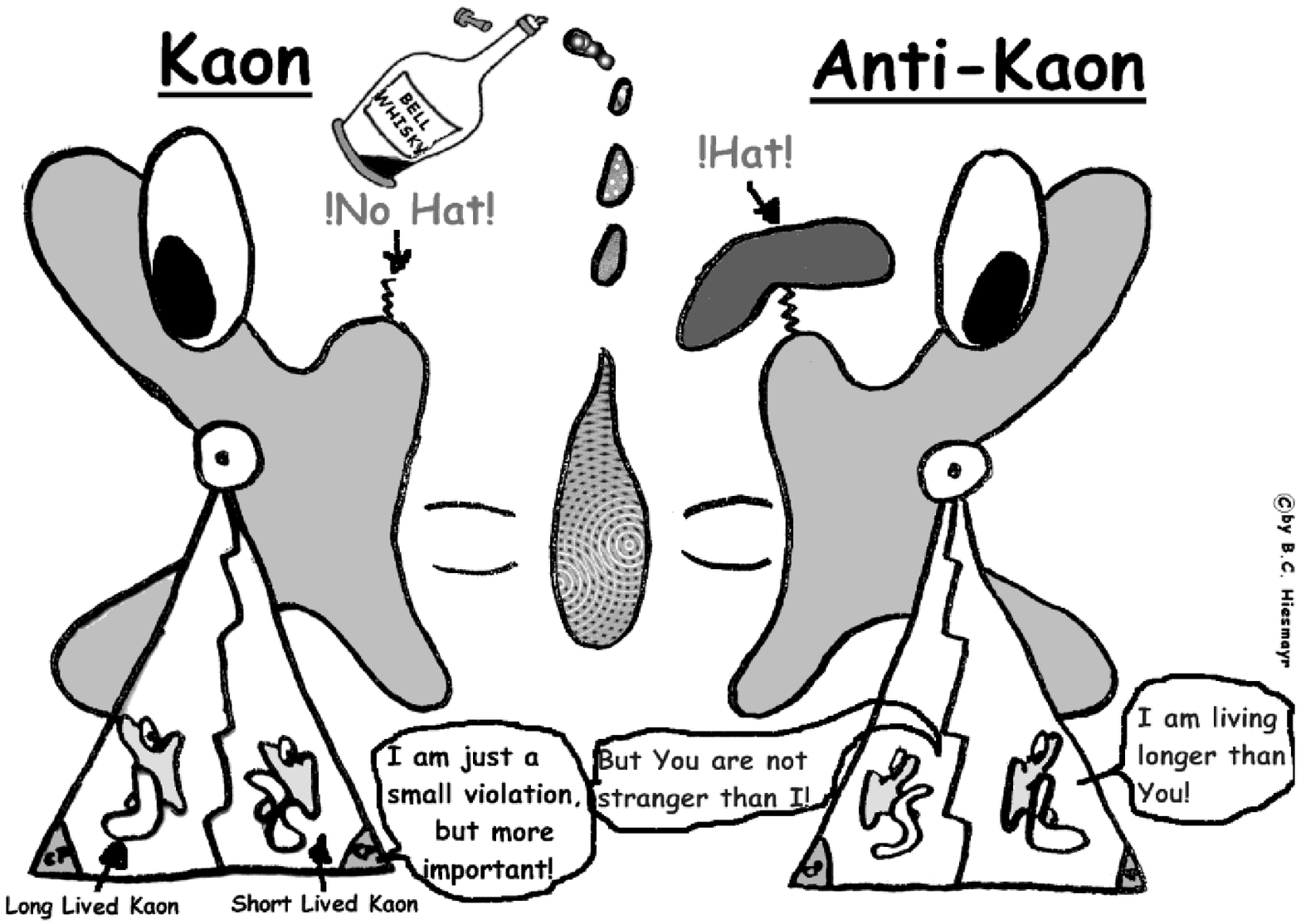}}
\end{titlepage}

\newpage

\vspace{4cm}
\begin{center}
\textbf{Front page picture}
\end{center}

This shows the production of two entangled neutral kaons. The hat symbolizes the strangeness
eigenstate, thus being a particle or an antiparticle. The tie symbolize the superposition
of a strangeness eigenstate of two mass eigenstates, the short lived one and the long lived one.
The little triangles in the ties stand for the $CP$ violation, which occurs in this
system.
The Whisky bottle is copied from \cite{Bertlmann}, which was a response of the
article of John S. Bell ``Bertlmann's socks and the nature of reality''
\cite{BellBertlmann}.

\vspace{4cm}

\begin{center}
\textbf{Acknowledgement}
\end{center}

The authors want to thank for the support of the Austrian FWF, project P14143-PHY,
and the Austrian-Czech Republic Scientific Collaboration, project KONTAKT 2001-11.

\newpage
\section{Introduction}


As it is well known John Bell's questioning orthodox quantum mechanics
was just his ``hobby'', and it is this ``hobby'' John Bell is most famous for
(Refs.\cite{bell2,bell}).
Though he was an authority at his working place CERN, with his
``hobby'' he was rather isolated there. And it took about 30 years
that Bell inequalities are now investigated in the particle physics community.
His broad knowledge and his deep understanding of physics were very impressive.
He wrote many classical papers in different areas, as you could witness at the Bell Conference
2000, held in Vienna in honour of him, or as you can witness by the book you hold
in your hands. One of his classical papers written 1965 together with Jack
Steinberger \cite{BellSteinberger} was about $CP$ violation of the neutral kaon system, and
curiously, nowadays precisely this system yields an opportunity to
investigate Bell inequalities
in massive systems. Compared to photons kaons are decaying and have $CP$
violation, which gives new features to the original EPR-paradox.

\section{The Bell-CHSH inequality for Photons and for
Kaons}\label{chshinequalityforphotonsandforkaons}

\vspace{0.2cm}
\begin{center}\begin{tabular}{||l||}
\emph{Before we explain the formalism of the neutral kaons, we compare the
Bell-Clauser}\\
\emph{-Horne-Shimony-Holt inequality for photons with the one for neutral kaons. We}\\
\emph{emphasize why
the neutral kaon system can be considered to have an EPR-like cor-}\\
\emph{relation, but show on the other hand already the differences.}
\end{tabular}\end{center}
\vspace{0.2cm}

The neutral kaon and its anti-particle can be distinguished by the strangeness
number $S$. This quantum number was introduced by Gell-Mann and Nishijima in 1953 to
solve the strange behaviour of these particles. They are produced nearly as often
as pions, but on the other hand they live long enough to travel a measurable
distance-- about several centimeter. This new quantum number is conserved by the
strong interaction, but violated by weak interaction, responsible for the decay of
the kaons. For this section it is important to know that due to the strong
interaction we can distinguish between the particle $K^0$, having $S=1$, and its anti-particle
$\bar{K^0}$, having $S=-1$.
A further nice feature of that neutral kaon system is that one of the four Bell-states,
the antisymmetric spin-singlet state or the polarisation $|\psi^-\rangle\sim|H\rangle|V\rangle
-|V\rangle|H\rangle$ state, can be produced:
\begin{flushleft}
\begin{tabular}{|c|}
\hline
$|\psi^{photon}\rangle=\frac{1}{\sqrt{2}}\biggl\lbrace|$\includegraphics[width=5.545mm,height=9.525mm]{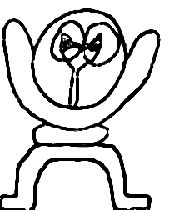}
$\rangle_l\otimes|$\includegraphics[width=5.545mm,height=9.525mm]{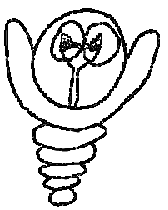}$\rangle_r-|$
\includegraphics[width=5.545mm,height=9.525mm]{vphoton5.eps}$\rangle_l\otimes |$
\includegraphics[width=5.545mm,height=9.525mm]{hphoton4.eps}$\rangle_r\biggr\rbrace$\\
\hline
\end{tabular}
\end{flushleft}
\begin{flushright}
\begin{tabular}{|c|}
\hline
$|\psi^{kaon}\rangle=\frac{1}{\sqrt{2}}\biggl\lbrace|$\includegraphics[width=5.545mm,height=9.525mm]{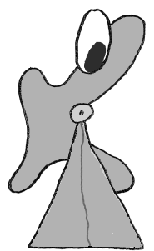}
$\rangle_l\otimes|$\includegraphics[width=5.545mm,height=9.525mm]{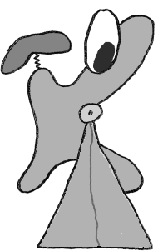}$\rangle_r-|$
\includegraphics[width=5.545mm,height=9.525mm]{akaon2.eps}$\rangle_l\otimes |$
\includegraphics[width=5.545mm,height=9.525mm]{kaon2.eps}$\rangle_r\biggr\rbrace$\\
\hline
\end{tabular}
\end{flushright}
Starting form that initial state for the photons we want to derive the probability of
detecting on the left $(l)$ side a photon behind a linear polarisator, whose optical axis is
turned around the angle $\alpha$, and on the right $(r)$ side a photon, whose
optical axis is turned around the angle $\beta$. For the neutral kaons we then
similarly have
to derive the probability of finding a neutral kaon $K^0$ on the
left side at time $t_l$ and on the right side also a $K^0$ but at time
$t_r$.\\
\\
\begin{small}
\begin{tabular}{|c l ||l c|}
\hline
\includegraphics[width=5.545mm,height=9.525mm]{hphoton4.eps}&
$P(\alpha,\beta)=\frac{1}{4}\big(1-\cos(2(\beta-\alpha))\big)
$&$P(K^0 t_l,K^0 t_r)=\frac{1}{8}\{e^{-\gamma_S t_l-\gamma_L
t_r}+e^{-\gamma_L t_l-\gamma_S t_r}$&\\
& &$\qquad\qquad\qquad - 2 \cos(\Delta m (t_l-t_r))\cdot e^{-\gamma (t_l+t_r)}\}$
&\includegraphics[width=5.545mm,height=9.525mm]{kaon2.eps}\\
\hline
\end{tabular}
\end{small}

Where $\gamma_S$ and $\gamma_L$ are decay width of the two mass eigenstates
($\gamma=\frac{1}{2} (\gamma_S+\gamma_L)$) and $\Delta m=m_L-m_S$
is the mass difference of these states. We see immediately that if we forbid the
kaons to decay we get the following probability:

\begin{small}
\begin{tabular}{|c l ||l c|}
\hline
\includegraphics[width=5.545mm,height=9.525mm]{vphoton5.eps}&
$P(\alpha,\beta)=\frac{1}{4}\big(1-\cos(2(\beta-\alpha))\big)
\quad$&NO Decays allowed: $\gamma_S=\gamma_L=0$&\\
& &$\vphantom{\biggr\rbrace}$\quad$P(K^0 t_l,K^0 t_r)=\frac{1}{4}\{1-\cos(\Delta m (t_l-t_r))\}$
&\includegraphics[width=5.545mm,height=9.525mm]{akaon2.eps}\\
\hline
\end{tabular}
\end{small}

Thus the probability of finding a $K^0$ on each side at the same time $t_l=t_r$ is zero,
whereas the probability of finding a $K^0$ on one side and $\bar K^0$ on the other side is
one, completely analog to the photon case if we choose $\alpha=\beta$. This is the reason why
we can speak of an EPR-like correlation of this massive system, but note that it is only
true for equal times on the left and the right side.\\
\\
\textbf{So we learn that the time difference $\Delta t=t_l-t_r$ in the kaon system plays
a similar role as the angle difference $\phi_{\alpha,\beta}=\beta-\alpha$ in the photon system. And
for $\Delta t=0$ we have this peculiar strong EPR-correlation, finding a kaon on
the right side and then a measurement at the same time on the other side gives us for sure an
anti-kaon (of course only if this kaon didn't yet decay).}\\

We have seen the similarities of the photon and neutral kaon system. So now the
natural question arises:\\
\textbf{Can we find a Bell-CHSH inequality for the kaons
-- similar to the photons -- in order to distinguish also in particle physics between
a local realistic theory and the quantum theory?}\\

The derivation of the Bell-CHSH inequality based on a local realistic theory (LRT) in the
kaon case is quite similar; however, then to calculate the quantum mechanical
expectation value is not so straightforward, because one has to include the decay states.
We will discuss this in the Section \ref{generalizedBI}. Here only the
result and its interesting implications are presented.

\begin{tabular}{l l}
\includegraphics[width=5.545mm,height=9.525mm]{hphoton4.eps}&
$\vphantom{\biggr\rbrace}$The CHSH inequality for the photon system is \cite{bell3,CHSH}\\
&$\vphantom{\biggr\rbrace}$$S^{photon}(\phi_{\alpha,\beta},\phi_{\alpha,\gamma},\phi_{\delta,\beta})=$\\
&$\qquad|\cos(\phi_{\alpha,\beta})-\cos(\phi_{\alpha,\gamma})|+|\cos(\phi_{\delta,\beta})+
\cos(-\phi_{\alpha,\beta}+\phi_{\alpha,\gamma}+\phi_{\delta,\beta})|\;\leq\;2$\\
&$\vphantom{\biggr\rbrace}$with $\phi_{\alpha,\beta}=2(\beta-\alpha)$.\\
\includegraphics[width=5.545mm,height=9.525mm]{akaon2.eps}&
The CHSH inequality for the kaon system is \cite{ghirardi91}\\
&$\vphantom{\biggr\rbrace}$$S^{kaon}(t_a,\phi_{a,b},\phi_{a,c},\phi_{d,b})=$\\
&$\qquad|\cos(\phi_{a,b})\cdot e^{-\gamma (t_a+t_b)}-
\cos(\phi_{a,c})\cdot e^{-\gamma (t_a+t_c)}|$\\
&$\qquad\qquad+|\cos(\phi_{d,b})\cdot e^{-\gamma (t_d+t_b)}+
\cos(-\phi_{a,b}+\phi_{a,c}+\phi_{d,b})\cdot e^{-\gamma (t_d+t_c)}|\;\leq\;2$\\
&$\vphantom{\biggr\rbrace}$with $\phi_{a,b}=\Delta m \;(t_b-t_a)$.
\end{tabular}\label{ghirardichoice}

Now lets analyze these two $S$-functions. Note that in the kaon case we actually
have 4 free parameters to choose. To see if the left hand side of the
inequality, respectively the $S$-function, gets bigger than 2, all we have to do is
to find the maximums. We carry coals to Newcastle, if we remind you that maximal violation
of the $S^{photon}$-function is $2 \sqrt{2}$. But what about the kaon system? The result for the
following choice of the parameters is
\begin{center}
\begin{tabular}{|c l||l c|}
\hline
\includegraphics[width=5.545mm,height=9.525mm]{hphoton4.eps}&$\vphantom{\biggr\rbrace}$$S^{photon}(\frac{3
\pi}{4},\frac{\pi}{4},\frac{\pi}{4})=2.828$&$S^{kaon}(0,\frac{3
\pi}{4},\frac{\pi}{4},\frac{\pi}{4})=0.426$
&\includegraphics[width=5.545mm,height=9.525mm]{kaon2.eps}\\
\hline
\end{tabular}\; .
\end{center}

Thus in this case we do not get any violation (larger $t_a$'s make the result even
worse). As the $S^{kaon}$-function is too complicated to analyze
analytically one has to handle it numerically \cite{ghirardi91,trixi}. One finds the highest
value for the following choices

\begin{center}
\begin{tabular}{|c l||l c|}
\hline
\includegraphics[width=5.545mm,height=9.525mm]{vphoton5.eps}&$\vphantom{\biggr\rbrace}$$S^{photon}(\frac{3
\pi}{4},\frac{\pi}{2},0)=2.414$&$S^{kaon}(0,\frac{3
\pi}{4},\frac{\pi}{2},0)=1.362$
&\includegraphics[width=5.545mm,height=9.525mm]{akaon2.eps}\\
\hline
\end{tabular}\; .
\end{center}

But this is still not bigger than 2! There is no choice of the four free parameters
which lets the S-function $S^{kaon}$
get bigger than 2, thus there is no way to distinguish between a local realistic
theory and quantum theory. It turns out, because the quantity\footnote{The quantity $x$ expresses the interplay of the
strangeness oscillation $\Delta m$ and the decay constant $\gamma_S$, see Section
\ref{aneutralkaonintroducesitself}.} $x=\frac{2 \Delta m}{\gamma_S}$
is about $1$ in the kaon system and not about a factor $4.3$ bigger, there is \emph{no} violation of
the CHSH inequality possible, already for theoretical reasons.\\
\\
\textbf{But is there really no way to distinguish between a LRT and QM?}\\
\\
As we show in this paper that there is another way to get an answer of that
tricky question, namely through a decoherence approach. In Section \ref{decoherence} we will work
out that approach in detail. There exist also different Bell inequalities for
neutral kaons using other properties of the neutral kaon system, for instance the
$CP$-violation, this we work out
in Section \ref{generalizedBI}. Last but not least we connect both approaches in
Section \ref{connection}. However to understand the following sections
additional information of the kaons and their properties is needed, so the next Section
represents such an overview.

\section{A neutral kaon introduces itself}\label{aneutralkaonintroducesitself}

\begin{tabular}{||l||}
\emph{Here we are going to learn something about the strange particles we have
been}\\
\emph{talking about.}
\end{tabular}
\vspace{0.5cm}

There exist to the two charged kaons $K^\pm$
with $S=\pm1$ two neutral kaons $K^0$ and $\bar K^0$, which form the isospin doublets
\renewcommand{\arraystretch}{1}
\begin{center}
\begin{tabular}{|c|c c c|}
\hline
\vphantom{$\frac{1^A}{2_A}$} & &$I_3$&\\
\vphantom{$\frac{1^A}{2_A}$}$S$&$+\frac{1}{2}$&&$-\frac{1}{2}$\\
\hline
\vphantom{$\frac{1^A}{2_A}$}$+1$&$K^+$&&$K^0$\\
\vphantom{$\frac{1^A}{2_A}$}$-1$&$\bar K^0$&&$K^-$\\
\hline
\end{tabular}$\;$.
\end{center}
Thus the neutral kaon is not its own anti-particle, it can be distinguished between the
particle and anti-particle through the strong interaction.

If one studies the strong interactions of the neutral kaons, one finds that $K^0$ and
$\bar K^0$ are pseudoscalar particles, hence the parity operator $P$ acts on the
neutral kaon as:
\begin{eqnarray}
P\; | K^0\rangle&=&-\;| K^0\rangle\nonumber\\
P\; | \bar K^0\rangle&=&-\;| \bar K^0\rangle.
\end{eqnarray}
The charge conjugation $C$ transforms a neutral kaon $K^0$ to its anti-particle $\bar
K^0$, the phases can be defined as follows:
\begin{eqnarray}
C\; | K^0\rangle&=&| \bar K^0\rangle\nonumber\\
C\; | \bar K^0\rangle&=&| K^0\rangle.
\end{eqnarray}
For the product $CP$ one has:
\begin{eqnarray}\label{odd}
CP \;| K^0\rangle &=& -\;| \bar K^0\rangle\nonumber\\
CP \;| \bar K^0\rangle &=& -\;| K^0\rangle.
\end{eqnarray}
The decay of the K-meson is a weak process. We know that the weak interaction neither
respect the strangeness $S$ nor the parity $P$ nor the charge conjugation $C$. In the
most cases the neutral kaons decay into two pions. Here can $K^0$ as well as $\bar K^0$
produce the same final state:
\begin{eqnarray}
& &K^0\;\;\Rightarrow\;\;\pi^+ \pi^-,\;\pi^0 \pi^0,\nonumber\\
& &\bar K^0\;\;\Rightarrow\;\;\pi^+ \pi^-,\;\pi^0 \pi^0.
\end{eqnarray}

\textbf{From that moment on it was clear that the weak interaction can induce in higher order
transitions between $K^0$ and $\bar K^0$!}\\
\\
This has as a result that the decay process cannot be considered separately, but has to
be handled as a two-state-system $K^0-\bar K^0$. A formalism for the decay mechanism of
an unstable state which is degenerate with one or more other states is called the
Wigner-Weisskopf approximation (1930) (see for example Refs.\cite{Nachtmann,Kabir}).

We deal with an effective Schr\"odinger equation
\begin{eqnarray}
i\frac{\partial}{\partial t}\;| \psi(t)\rangle=\mathcal{H}\;| \psi(t)\rangle\;,
\end{eqnarray}
where $\mathcal{H}$ is an operator in the two dimensional space of the neutral kaons and is in
general not-hermitian. One can show that this operator can be separated in a hermitian and
not-hermitian part. These matrices are the generalization of the mass and decay-width of a
decaying particle without state-mixing.

Now we consider the eigenvalue problem of $\mathcal{H}$. The eigenvectors will be
called $| K_S\rangle$, $| K_L\rangle$ with the eigenvalue $\lambda_S$, $\lambda_L$:
\begin{eqnarray}
\mathcal{H}\;| K_S\rangle&=&\lambda_S\; | K_S\rangle,\nonumber\\
\mathcal{H}\;| K_L\rangle&=&\lambda_L\; | K_L\rangle.
\end{eqnarray}
Because $\mathcal{H}$ is not-hermitian, so neither the eigenvalues are real nor
the eigenstates are necessarily orthogonal, one denotes
\begin{eqnarray}
\lambda_S&=& m_S-\frac{i}{2} \gamma_S,\nonumber\\
\lambda_L&=& m_L-\frac{i}{2} \gamma_L,
\end{eqnarray}
where $m_S, m_L, \gamma_S, \gamma_L$ are all real. The time evolution in this
basis is exponential
\begin{eqnarray}
|K_S(t)\rangle&=& e^{-i \lambda_S t}\;|K_S\rangle\qquad\qquad\textrm{with}\qquad
\lambda_S=m_S-\frac{i}{2}\gamma_S\nonumber\\
|K_L(t)\rangle&=& e^{-i \lambda_L t}\;|K_L\rangle\qquad\qquad\textrm{with}\qquad
\lambda_L=m_L-\frac{i}{2}\gamma_L\;.
\end{eqnarray}

The experimental values of these quantities are
\begin{eqnarray}
\tau_S&=&\frac{1}{\gamma_S}=(0.8935\pm0.0008)\cdot 10^{-10}s\nonumber\\
\tau_L&=&\frac{1}{\gamma_L}=(5.17\pm0.04)\cdot 10^{-8}s\nonumber\\
\Delta m&=&m_L-m_S=(0.5300\pm0.0012)\cdot10^{10} s^{-1}\;.
\end{eqnarray}

There exist two different mass-eigenstates, the short lived eigenstate $|K_S\rangle$
and the long lived eigenstate $|K_L\rangle$ which lives about $600$ times longer
than the short lived kaon $|K_S\rangle$. The mass difference $\Delta m$ is responsible for the
strangeness oscillation, thus it happens that an initial kaon transforms with a certain probability
into an anti-kaon.

In (\ref{odd}) we have seen that both the kaon and the anti-kaon are in an odd $CP$
eigenstate. Now it is easy to construct the $CP$ eigenstates
\begin{eqnarray}
|K^0_1\rangle&=&\frac{1}{\sqrt{2}}\{|K^0\rangle-|\bar K^0\rangle\}\nonumber\\
|K^0_2\rangle&=&\frac{1}{\sqrt{2}}\{|K^0\rangle+|\bar K^0\rangle\}\;.
\end{eqnarray}

In the decay processes it was observed that the short lived kaon decayed
into 2 pions and the long lived kaon decayed into 3 pions. Two pions are in a $CP=+1$
state and  a three pion state has $CP=-1$. So it was naturally to identify
the short lived state $|K_S\rangle$ with the $CP=+1$ state $|K_1^0\rangle$ and the
long lived state $|K_L\rangle$ with $|K_2^0\rangle$.

However, in 1964 the famous experiment by Christensen, Cronin, Fitch and Turlay could
demonstrate that the long lived kaon also decays into two pions.

\begin{center}\textbf{This means $CP$-symmetry is broken!}\end{center}

Anyway the value of the $CP$ symmetry, the $CP$ parameter
$\varepsilon$, is small, about $10^{-3}$.

For the neutral kaons this means that we have the following physically important
quasi-spin eigenstates\footnote{The generalized definition of a quasi-spin eigenstate is to be an arbitrary
superposition of the strangeness eigenstates.}
\begin{eqnarray}\label{definitioneigenstates}
|K^0\rangle\;\hphantom{=\frac{1}{N}\{p |K^0\rangle-q |\bar K^0\rangle\}}\qquad&
|\bar K^0\rangle\;\hphantom{=\frac{1}{N}\{p |K^0\rangle-q |\bar K^0\rangle\}}\;\;\nonumber\\
|K_S\rangle=\frac{1}{N}\{p |K^0\rangle-q |\bar K^0\rangle\}\qquad&
|K_L\rangle=\frac{1}{N}\{p |K^0\rangle+q |\bar K^0\rangle\}\;\;\nonumber\\
|K^0_1\rangle=\frac{1}{\sqrt{2}}\{\;|K^0\rangle-\;|\bar K^0\rangle\}\qquad&
|K^0_2\rangle=\frac{1}{\sqrt{2}}\{\;|K^0\rangle+\;|\bar K^0\rangle\}\;.
\end{eqnarray}

with $p=1+\varepsilon, q=1-\varepsilon$ and $N^2=|p|^2+|q|^2$.

\section{The experiment at CERN and possible decoherence}\label{decoherence}

\begin{tabular}{||l||}
\emph{In this section we introduce the CPLEAR experiment  performed at
CERN.}\\
\emph{We describe the hypothesis that spontaneous decoherence of the wave function
}\\
\emph{takes places and we illustrate its consequences. With the help of the data of the}\\
\emph{CPLEAR experiment we estimate the possible values for the ``decoherence }\\
\emph{parameter''
$\zeta$.}
\end{tabular}

\vspace{0.5cm}

This experiment \cite{CPLEAR-EPR} was performed in 1998. The $K^0 \bar K^0$ pairs were produced in
a\\
$J^{PC}=1^{--}$ state -- the one corresponding to the $|\psi^-\rangle$ state of the
photons -- by proton-antiproton annihilation. So the initial state of that
production is
\begin{eqnarray}\label{psik0bk0}
|\psi(t=0)\rangle=\frac{1}{\sqrt{2}}\big\lbrace | K^0\rangle_l\otimes |\bar K^0\rangle_r-| \bar K^0\rangle_l\otimes
|K^0\rangle_r\big\rbrace
\end{eqnarray}

and with the knowledge of the previous Section \ref{aneutralkaonintroducesitself} we
can rewrite this state in the mass eigenstate basis
\begin{eqnarray}\label{psikskl}
|\psi(t=0)\rangle=\frac{1}{2 \sqrt{2} p q}\big\lbrace | K_S\rangle_r\otimes |K_L\rangle_l-|K_L\rangle_r\otimes
|K_S\rangle_l\big\rbrace\;.
\end{eqnarray}

The CPLEAR group constructed two different setups, shown in Fig.1.

\begin{multicols}{2}
\center{\includegraphics[width=160pt, height=140pt]{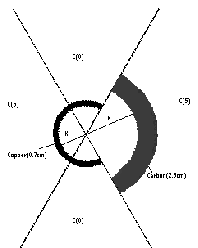}}\\
\begin{flushleft}
\footnotesize{Fig.1: A section through the experimental
construction of the CPLEAR experiment at CERN in 1998. The $C(0)$
region shows the EPR-like configuration. Both kaons have nearly
equal proper times when they interact with the copper absorber
which is about $2 cm$ away from the creation point. The $C(5)$
configuration shows a proper time difference of $\Delta t=5 cm$;
the carbon absorber is about $7 cm$ away.}
\end{flushleft}
\end{multicols}

The first one is called the $C(0)$ configuration, here each kaon travels $2 cm$,
thus the time difference $\Delta t$ of the two flying paths is zero. This is the EPR-like
case, detecting on the right side a kaon with $S=+1$ implies that on the left
side there is no such kaon with $S=+1$. The second configuration is called $C(5)$,
because now the time difference is about $5 cm$, respectively one kaon is detected
after a flying path of $2 cm$ and the second one after a flying path of $7 cm$.

What the CPLEAR group measured, was the difference between the like-strangeness
events and the unlike-strangeness events, i.e., how often
they detected the two kaons with $S=1$ or with $S=-1$ (like-strangeness events) and how
often they detected one kaon with $S=1$ and the other one with $S=-1$
(unlike-strangeness events).



%
This quantity is normed by the sum of these two probabilities and is called
asymmetry term:
\begin{eqnarray}\label{asymmetry}
A^{QM}(t_l,t_r)=
\frac{P_{unlike}(t_l,t_r)-P_{like}(t_l,t_r)}{P_{unlike}(t_l,t_r)+P_{like}(t_l,t_r)}.
\end{eqnarray}

In quantum mechanics such a like-strangeness probability $P_{like}(t_l,t_r)$\footnote{The
$CP$-violation is neglected in the whole section, because it gives only
corrections of the order $10^{-3}$ and this is far away form the experimental accuracy.}
is calculated in a straight forward way:
\begin{eqnarray}\label{plike}
P_{like}(t_l,t_r)&=&||\langle K^0|_l \otimes \langle
K^0|_r\;|\psi(t_l,t_r)\rangle||^2=||\langle \bar K^0|_l \otimes \langle
\bar K^0|_r\;|\psi(t_l,t_r)\rangle||^2=\nonumber\\
&=&\frac{1}{8}\biggl\lbrace e^{-\gamma_S t_l-\gamma_L t_r}+e^{-\gamma_L t_l-\gamma_S
t_r}-2 \cos(\Delta m \Delta t)\cdot e^{-\gamma (t_l+t_r)}\biggr\rbrace
\end{eqnarray}

with $\Delta t=t_l-t_r$ (for a definition of $\psi(t_l,t_r)$ see Eq.(\ref{unitarypsi})).
Measuring the same strangeness on both
sides at the same time $t_l=t_r$ has probability zero. Similarly the unlike-strangeness
probability yields the result
\begin{eqnarray}
P_{unlike}(t_l,t_r)&=&||\langle K^0|_l \otimes \langle
\bar K^0|_r\;|\psi(t_l,t_r)\rangle||^2=||\langle \bar K^0|_l \otimes \langle
K^0|_r\;|\psi(t_l,t_r)\rangle||^2=\nonumber\\
&=&\frac{1}{8}\biggl\lbrace e^{-\gamma_S t_l-\gamma_L t_r}+e^{-\gamma_L t_l-\gamma_S
t_r}+2 \cos(\Delta m \Delta t)\cdot e^{-\gamma (t_l+t_r)}\biggr\rbrace\;.
\end{eqnarray}

We note that the interference term, the strangeness oscillation, changed its sign,
so for equal time measurements this unlike-strangeness probability maximizes. Inserting the two
probabilities in our asymmetry term (\ref{asymmetry}) gives
\begin{eqnarray}
A^{QM}(\Delta t)=\frac{\cos(\Delta m \Delta t)}{\cosh(\frac{1}{2}\Delta \gamma \Delta
t)}\qquad\qquad \textrm{with}\quad \Delta \gamma=(\gamma_L-\gamma_S).
\end{eqnarray}

The asymmetry term depends only on the time difference $\Delta t$ and is direct proportional
to the interference term, the strangeness oscillation. This was the quantity which
was tested in the CPLEAR experiment. The experimental results for the two
configurations of the asymmetry term are the following
\begin{center}
\begin{tabular}{|r|c|c|}
\hline
&Experiment& Theory (corrected)\\
\hline
$C(0)$&$0.81\pm0.17$&$0.93$\\
$C(5)$&$0.48\pm0.12$&$0.56$\\
\hline
\end{tabular}
\end{center}

and they are compared with the according to the experimental configuration corrected theoretical
values.
The experimental values agree within one standard deviation with quantum theory. And the
peculiar quantum entanglement seems to be confirmed.

\begin{center}
\textbf{But is it really that way?}
\end{center}

What is, if we image that the initial state (\ref{psikskl}) immediately
after its creation collapses spontaneously into its components, e.g.,
\begin{eqnarray}\label{ksklfactorization}
& &|\psi\rangle\quad\Longrightarrow\quad
|K_S\rangle_l\otimes|K_L\rangle_r\qquad\textrm{in half of the cases}\nonumber\\
& &|\psi\rangle\quad\Longrightarrow\quad
|K_L\rangle_l\otimes|K_S\rangle_r\qquad\textrm{in the other half.}
\end{eqnarray}

This factorization of the wave function is called Furry's hypothesis\footnote{Actually it should
be called
Schr\"odinger's hypothesis, because first he stated it already one year earlier \cite{Schroedinger}
and second he remarked that it could really happen (to be reread in
\cite{ClauserShimony}).} \cite{Furry}.

We will now modify the calculation of the probabilities in the way that we have on
one hand the quantum mechanical probability and on the other hand Furry's hypothesis
or spontaneous factorization of the initial wave function. For this we again look at the
derivation of the quantum mechanical like-strangeness probability (\ref{plike}) and
modify it in the following way (Refs.\cite{BGH,trixi}):
\begin{eqnarray}\label{plikezeta}
\lefteqn{P_{like}(t_l,t_r)=||\langle K^0|_l \otimes \langle
K^0|_r\;|\psi(t_l,t_r)\rangle||^2\quad\longrightarrow\quad P^\zeta_{like}(t_l,t_r)}\nonumber\\
&=&\frac{1}{2}\biggl\lbrace e^{-\gamma_S t_l-\gamma_L t_r} |\langle K^0|K_S\rangle_l|^2\;|\langle
K^0|K_L\rangle_r|^2+
e^{-\gamma_L t_l-\gamma_S t_r}|\langle K^0|K_L\rangle_l|^2\; |\langle K^0|K_S\rangle_r|^2
\nonumber\\
& &-2 \underbrace{(1-\zeta)}\; Re\{\langle K^0|K_S\rangle_l^*\;\langle K^0|K_L\rangle_r^*\langle K^0|K_L\rangle_l\;
\langle K^0|K_S\rangle_r e^{+i \Delta m \Delta t}\}\cdot e^{-\gamma
(t_l+t_r)}\biggr\rbrace\nonumber\\
& &\quad\textrm{modification}\nonumber\\
&=&\frac{1}{8}\biggl\lbrace e^{-\gamma_S t_l-\gamma_L t_r}+e^{-\gamma_L t_l-\gamma_S
t_r}-2 \underbrace{(1-\zeta)}\cos(\Delta m \Delta t)\cdot e^{-\gamma
(t_l+t_r)}\biggr\rbrace.\nonumber\\
& &\hphantom{\frac{1}{8}\biggl\lbrace e^{-\gamma_S t_l-\gamma_L t_r}+e^{-\gamma_L t_l-\gamma_S
t_r}}\quad\textrm{modification}
\end{eqnarray}

Thus we have multiplied the interference term of the decay amplitudes by a factor
$(1-\zeta)$. For $\zeta=0$ we have the quantum mechanical expression for the like
strangeness probability, but if $\zeta=1$, the quantum mechanical
interference term of the decay amplitudes vanish, thus the wave function factorizes as shown in
(\ref{ksklfactorization}). We actually found a way to have both the quantum
mechanical result on one hand and the factorization on the other hand, just
depending on the choice of the decoherence parameter $\zeta$.

All we have now to do, is to recalculate the asymmetry term (\ref{asymmetry}) with
this simple modification and compare it with the measurement results of the CPLEAR
experiment.

\begin{multicols}{2}
The asymmetry term reads
\begin{eqnarray}\label{aymmetryksklzeta}
A_\zeta(\Delta t)=A^{QM}(\Delta t)\cdot (1-\zeta),
\end{eqnarray}
thus is only linearly effected by our modification. Obviously for $\zeta$ equal zero, we have the
quantum mechanical result, but for $\zeta$ equal one, the asymmetry term (\ref{aymmetryksklzeta})
vanish for all time differences.
We have already considered the results of the CPLEAR experiment and have seen that they did not measure
values equal to zero, however, if we compare both experimental values and their corresponding
uncertainties
with the modified asymmetry term we get the following fit result\footnote{Of course
fitting the experimental results with such a modified theory requires more
considerations which have been made and explained in \cite{BGH,trixi}, however, the correct
result with a Confidential Level (C.L.) of $97\%$ is given here.}
\begin{eqnarray}\label{fitresultkskl}
\bar \zeta=0.13^{+0.16}_{-0.15}\;.
\end{eqnarray}
The result is also printed in Fig.2.\\

\flushleft{\includegraphics[width=210pt, height=147pt]{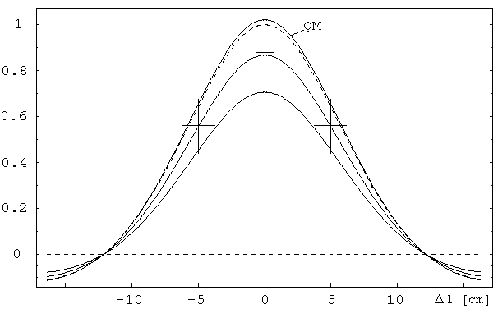}
\footnotesize{Fig.2: The asymmetry (\ref{aymmetryksklzeta}) as a function of the difference in
the distances traveled by the kaons to the points where their strangeness is
mea-\\ sured. The dashed curve corresponds to QM with the decoherence parameter
$\zeta=0$, whereas the solid curves correspond to the values of $\zeta$ obtained by
the fit (\ref{fitresultkskl}) to the CPLEAR data. The two data points represented by
the crosses have been taken from Ref.\cite{CPLEAR-EPR}. The horizontal dashed line
indicates the zero asymmetry for $\zeta=1$, the consequence of Furry's hypothesis
with respect to the $K_S K_L$ basis.}}
\end{multicols}

Now we could conclude that no such factorization is possible due to the data of
the CPLEAR experiment.

\begin{center}
\textbf{But is this really true?}\\
\end{center}

Lets go back to Eq.(\ref{ksklfactorization}), why should it actually factorize this
way, why not so:
\begin{eqnarray}\label{k0bk0factorization}
& &|\psi\rangle\quad\Longrightarrow\quad
|K^0\rangle_l\otimes|\bar K^0\rangle_r\qquad\textrm{in half of the cases}\nonumber\\
& &|\psi\rangle\quad\Longrightarrow\quad
|\bar K^0\rangle_l\otimes|K^0\rangle_r\qquad\textrm{in the other half.}
\end{eqnarray}

In the quantum theory one has the free choice of the basis, so it is natural to
consider also such a case. For that we have to start from the initial state in
the $K^0\bar K^0$ basis choice (\ref{psik0bk0}) and then recalculate the modified
probabilities
\begin{eqnarray}\label{plikek0bk0}
P_{like}(t_l,t_r)&=&||\langle K^0|_l \otimes \langle
K^0|_r\;|\psi(t_l,t_r)\rangle||^2\quad\longrightarrow\nonumber\\
P^\zeta_{like}(t_l,t_r)&=&\frac{1}{2}\biggl\lbrace
|\langle K^0|K^0(t_l)\rangle_l|^2\;|\langle K^0|\bar K^0(t_r)\rangle_r|^2+
|\langle K^0|\bar K^0(t_l)\rangle_l|^2\;|\langle K^0|K^0(t_r)\rangle_r|^2\nonumber\\
&&-2\underbrace{(1-\zeta)} Re\{
\langle K^0|K^0(t_l)\rangle_l^*\;\langle K^0|\bar K^0(t_r)\rangle_r^*
\langle K^0|\bar K^0(t_l)\rangle_l\;\langle
K^0|K^0(t_r)\rangle_r\}\biggr\rbrace\nonumber\\
& &\quad\textrm{modification}
\end{eqnarray}

It is clear that the decay amplitudes starting with the $K_S K_L$ basis differ from
the ones in the $K^0 \bar K^0$ basis, since the interference term of (\ref{plikezeta})
differs from the one in (\ref{plikek0bk0}). Or put simply, the factorization in the $K_S K_L$
basis (\ref{ksklfactorization}) is not equal to the factorization in the $K^0\bar K^0$
(\ref{k0bk0factorization}). In this sense the decoherence approach is basis depend.

Now we can derive the asymmetry term with the decoherence parameter $\zeta$ in the
$K^0 \bar K^0$ basis choice and we find
\begin{eqnarray}
A^{K^0\bar K^0}_\zeta(t_l,t_r)=\frac{\cos(\Delta m \Delta t)-\frac{1}{2}\zeta\bigl\lbrace
\cos(\Delta m \Delta t)-\cos(\Delta m (t_l+t_r))\bigr\rbrace}
{\cosh(\frac{1}{2}\Delta \gamma \Delta t)-\frac{1}{2}\zeta\bigl\lbrace\cosh(\frac{1}{2}\Delta \gamma \Delta
t)-\cosh(\frac{1}{2}\Delta \gamma (t_l+t_r)\bigr\rbrace}.
\end{eqnarray}
This asymmetry term is a little more complicated, it depends not only on the time
difference $\Delta t$, but also on the sum of the two involved times, and, moreover,
the dependence on $\zeta$ is not linear. Again we compare this
asymmetry term with the experimental data, the result\footnote{The C.L. is about
$67\%$, see \cite{BGH,trixi}} is
\begin{eqnarray}\label{fitresultk0bk0}
\bar \zeta\;\sim\;0.4\pm0.7.
\end{eqnarray}
But this is curious! Within one standard deviation both the quantum mechanical
result and Furry's hypothesis is included. In this basis choice we can't distinguish
between QM and a spontaneous factorization of the wave function!\\

\textbf{R\'{e}sum\'{e}}\\

\emph{We learned that Furry's hypothesis is basis dependent and so is our decoherence
approach. With help of the CPLEAR data we could restrict the degree of decoherence,
although an arbitraryness of the basis choice remains.}

\emph{However, what is essential, is
the existence of a basis where the $K^0\bar K^0$ system is far away from total
decoherence and the corresponding $\zeta$ is close to zero in agreement with QM. We
have seen the ``best basis'' in this respect is the $K_S K_L$ basis. So in this
sense we demonstrated the existence of interference effects of massive particles
over macroscopic distances (remember they were separated several centimeters).}\\
\\
Now naturally the question arises, how this decoherence parameter is really connected
to hidden variable theories. Another question arises, can we somehow use the other
properties of the neutral kaons, which have no similarities to the photon system,
and construct Bell inequalities. As curious as it is, the answer is yes, but
with some limitations, as we will see now.

\section{The generalized Bell inequality and unitary time
evolution}\label{generalizedBI}

\begin{tabular}{||l||}
\emph{Here we will derive a Bell-CHSH inequality based on locality, realism and
induction}\\
\emph{for different times (this is then the one analyzed in Section
\ref{chshinequalityforphotonsandforkaons}) but also vary the quasi}\\
\emph{spin eigenstates.
We analyze this generalized Bell inequality for different choices.}
\end{tabular}
\vspace{0.5cm}

When testing QM against LRT we find analogies but also important physical
differences between kaons and spin $1/2$ particles\footnote{We will now talk about the spin $1/2$
systems, because of the analogy to kaons; but the reader common with photons can
just image the photon system.} (for an alternative view
see Refs.\cite{Bramon, Gisin}).
\begin{enumerate}
\item While in the spin $1/2$ case one can test whether a system is in an
arbitrary state $\alpha |\Uparrow_n\rangle+
\beta |\Downarrow_n\rangle$, there is no analogous way to test
the arbitrary superposition
$\alpha |K^0\rangle+\beta |\bar K^0\rangle$. However, as in Ref.\cite{Uchiyama,
domenico}, we will assume that the mass and $CP$ eigenstates (\ref{definitioneigenstates}) can be
measured by a gedanken experiment.

\item While in the
spin $1/2$ case the direct product space $H^l_{spin} \otimes H^r_{spin}$ is sufficient to account for
all spin properties of the entangled system, this is \emph{not} true for the
neutral kaon case.

Indeed, we want to emphasize that due to the unitary time evolution of the states one has to
include the decay product spaces which are orthogonal to the product space $H^l_{kaon} \otimes
H^r_{kaon}$. This leads to additional terms
in the resulting Wigner-type inequality compared to other works
\cite{domenico,Benattiepsilonstrich,Gisin}
and compared to the photon case.
\end{enumerate}


In  the case of spin variables one can derive the common
Bell-CHSH inequality \cite{bell3, CHSH} for the averaged spin values along
arbitrary quantization directions $n$ and $m$. The analogue in the kaon case is the free
choice of the times at which
measurements are performed and in addition the freedom of choosing the
quasi-spin state of the meson, the strangeness eigenstate, the mass eigenstate or
the $CP$ eigenstate.

\begin{center}
\textbf{How to derive a Bell-CHSH inequality in a local realistic theory for the
neutral kaons?}
\end{center}

Bell's locality assumption requires then that the results at one side be completely
independent of the choice of the time and the choice of the quasi spin eigenstates of the
other side. To define the appropriate
correlation functions to be used in a Bell's inequality, we consider an observable
$O^l(k_n,t_a)$ on the left side, which gets the value $+1$ if the measurement at
time point $t_a$ gives the quasi spin $k_n$ and the value $-1$
if the quasi spin $k_n$ is not found. In terms of such an observable we can
define the correlation
function $O(k_n t_a; k_m t_b)$, which takes the value $+1$ both when a $k_n$ at $t_a$
and a $k_m$ at $t_b$ was detected or when no $k_n$ and no $k_m$ was detected. In the case that
only one of the desired quasi spin eigenstate has been found, no matter at which
side, the correlation function takes the value $-1$.

The locality assumption implies then that $O(k_n t_a, k_m t_b)$, in a specific
individual experiment, equals the product of $O^l(k_n, t_a)$ and $O^r(k_m, t_b)$:
\begin{eqnarray}
O(k_n t_a; k_m t_b)&=& O^l(k_n, t_a)\;\cdot\; O^r(k_m, t_b).
\end{eqnarray}

It implies
\begin{eqnarray}
|O(k_n t_a; k_m t_b)- O(k_n t_a; k_{m'} t_c)|\;+\;|O(k_{n'} t_d; k_{m'} t_c)+O(k_{n'} t_d;
k_m t_b)|\;= 2
\end{eqnarray}

with $k_n, k_m, k_{m'}$ and $k_{n'}$ being arbitrary quasi spin eigenstates of the meson
and $t_a, t_b, t_c$ and $t_d$ four different times.

Let us now consider a sequence of $N$ identical measurements, and let us denote by $O_i$
the value taken by $O$ in the $i$-th experiment. The average is given by
\begin{eqnarray}
M(k_n t_a; k_m t_b)&=&\frac{1}{N} \sum_{i=1}^N O_i(k_n t_a; k_m t_b)
\end{eqnarray}

and satisfies the Bell-CHSH inequality
\begin{eqnarray}\label{chsh-inequality}
& &| M(k_n t_a; k_m t_b)-M(k_n t_a; k_{m'} t_c)|+| M(k_{n'} t_d; k_{m'} t_c)+
M(k_{n'} t_d; k_m t_b)|\; \leq \nonumber\\
& &\frac{1}{N} \sum_{i=1}^N \big\lbrace
|O_i(k_n t_a; k_m t_b)- O_i(k_n t_a; k_{m'} t_c)|\;+\;|O_i(k_{n'} t_d; k_{m'} t_c)+O_i(k_{n'} t_d;
k_m t_b)|\;=\;2.\nonumber\\
\end{eqnarray}

Note, setting $M(k_n t_a; k_m t_b)$ equivalent to $M(\vec{a},\vec{b})$
Eq.(\ref{chsh-inequality}) reads exactly like the CHSH inequality for photons.\\

\textbf{How to derive the appropriate quantum probabilities for the neutral
kaons?}\\

As we have emphasized, a unitary time evolution for the neutral kaons is necessary.
Note, that in the photon system we have automatically unitarity. Such a unitary time
evolution for neutral kaons looks like
\begin{eqnarray}
U(t,0)|K_{S,L}\rangle=e^{-i \lambda_{S,L} t} |K_S\rangle+|\Omega_{S,L}(t)\rangle
\end{eqnarray}

where $|\Omega_{S,L}(t)\rangle$ describes the decay products of the neutral kaons and
operates
in a Hilbert space orthogonal to $H_{kaon}$.
Thus we operate in a complete Hilbert space, analogously to the photon case. The
time evolution of the initial state $|\psi(0)\rangle$ Eq.(\ref{psikskl}) is
(for details please see \cite{BH})
\begin{eqnarray}\label{unitarypsi}
|\psi(t_l,t_r)\rangle&=&U_l(t_l)\otimes U_r(t_r)\;|\psi(0)\rangle\nonumber\\
&=& (e^{-i \lambda_S t_l}|K_S\rangle_l+|\Omega_S(t_l)\rangle_l)\otimes
(e^{-i \lambda_L t_r}|K_L\rangle_r+|\Omega_L(t_r)\rangle_r)\nonumber\\
& &\quad\quad -
(e^{-i \lambda_L t_l}|K_L\rangle_l+|\Omega_L(t_l)\rangle_l)\otimes
(e^{-i \lambda_S t_r}|K_S\rangle_r+|\Omega_S(t_r)\rangle_r)\;.
\end{eqnarray}

Now the only thing we need to do to get, for example, the probability of finding the
a special quasi-spin state $|k_n\rangle$ on the left side at $t_l$ (a yes (Y) event) and
another quasi-spin state $|k_m\rangle$ on the right side at $t_r$ (Y event), is to apply the
corresponding projection operators to that state (\ref{unitarypsi}) and square it
\begin{eqnarray}
P_{k_n,k_m}(Y t_l, Y t_r)&=&||P_l(k_n)\otimes P_r(k_m)\;|\psi(t_l,t_r)\rangle||^2\;.
\end{eqnarray}

As well we can ask what is the probability of finding on the left side a $|k_n\rangle$
at $t_l$ and on the right side \textbf{no} $|k_m\rangle$ at $t_r$ (no (N) event). This means
for the experimenter that she has a detector for observing a $|k_m\rangle$ state and she
finds no such state at the time $t_r$
\begin{eqnarray}
P_{k_n,k_m}(Y t_l, N t_r)&=&||P_l(k_n)\otimes (1-P_r(k_m))\;|\psi(t_l,t_r)\rangle||^2.
\end{eqnarray}

In this way our quantum mechanical expectation value is given by
\begin{eqnarray}\label{qmmeanvalue}
& &M^{QM}(k_n t_a; k_m t_b)=\nonumber\\
& &P_{n,m}(Y t_a, Y t_b)+P_{n,m}(N t_a, N t_b)-P_{n,m}(Y t_a, N t_b)
-P_{n,m}(N t_a, Y t_b)\;.\nonumber\\
\end{eqnarray}

Further we can use that the sum of the probabilities of the results $(Y,Y)$,
$(N,N)$, $(Y,N)$ and $(N,Y)$ is unity for all times, so Eq.(\ref{qmmeanvalue}) can be
rewritten to
\begin{eqnarray}
M^{QM}(k_n t_a; k_m t_b)=-1 + 2 \big\lbrace P_{n,m}(Y t_a, Y t_b)+P_{n,m}(N t_a,N t_b) \big\rbrace.
\end{eqnarray}

Setting this expression into the Bell-CHSH inequality (\ref{chsh-inequality}) we get the
following inequality
\begin{eqnarray}\label{dieSgleichung}
\lefteqn{S^{kaon}(k_n t_a; k_m t_b; k_{n'} t_c; k_{m'} t_d)=}\nonumber\\
&=&|P_{n,m}(Y t_a, Y t_b)+P_{n,m}(N t_a, N t_b)-P_{n,n'}(Y t_a, Y t_c)-P_{n,n'}(N t_a, N
t_c)|\nonumber\\
& &+|-1+P_{m',m}(Y t_d, Y t_b)+P_{m',m}(N t_d, N t_b)\nonumber\\
& &\hphantom{+|-1+P_{m',m}(Y t_d, Y t_b)}
+P_{m',n'}(Y t_d, Y t_c)+P_{m',n'}(N t_d, N
t_c)|\;\leq\;1\;.\nonumber\\
\end{eqnarray}
This is our generalized Bell-CHSH inequality for the neutral kaons! Now we have a
lot of possibilities to choose the 8 parameters. Let us first choose it in the way
that
we get the Bell-CHSH inequality of Section \ref{chshinequalityforphotonsandforkaons}.
\\
\\
\textbf{1. The choice of the strangeness eigenstate}\\
\\
In this case we have to choose all quasi-spin eigenstates to be the same,
e.g., the anti-kaon\footnote{The choice of the anti-kaon has experimental reasons; this
strangeness state reacts with matter stronger than the one with $S=1$.} \cite{ghirardi91}
\begin{eqnarray}
k_n=k_m=k_{n'}=k_{m'}= \bar K^0.
\end{eqnarray}
Calculating now all involved probabilities and inserting them into (\ref{dieSgleichung}) and
the result\footnote{$CP$ violation is neglected and we have put $\gamma_L=0$.} is
\begin{eqnarray}\label{chshghirardietal}
& &|e^{-\frac{\gamma_S}{2} (t_a+t_b)} \, \cos(\Delta m (t_a-t_b))
- e^{-\frac{\gamma_S}{2} (t_a+t_c)} \, \cos(\Delta m (t_a-t_c))|\nonumber\\
& & + |e^{-\frac{\gamma_S}{2} (t_d+t_b)} \, \cos(\Delta m (t_d-t_b))
+ e^{-\frac{\gamma_S}{2} (t_d+t_c)} \, \cos(\Delta m (t_d-t_c))|\;\leq\;2\,
.\nonumber\\
\end{eqnarray}

Unfortunately, -- as discussed in Section \ref{chshinequalityforphotonsandforkaons} --
this inequality (\ref{chshghirardietal}) \emph{cannot}
be violated for any choice of the four (positive)
times $t_a, t_b, t_c, t_d$ due to the interplay between the kaon decay width and
the strangeness oscillation.\\
\\
\textbf{2. The choice sensitive to the $CP$ violating parameter $\varepsilon$}\\
\\
Now we will set all times equal to $t_a=t_b=t_c=t_d=0$ and further choose $n'=m'$; the
remaining states we choose in the following way
\begin{eqnarray}
|k_n\rangle\,&=&|K_S\rangle\nonumber\\
|k_m\rangle&=&|\bar K^0\rangle\nonumber\\
|k_{n'}\rangle&=&|K_1^0\rangle\;,
\end{eqnarray}

and if we denote the probabilities $P_{K_S,\bar K^0}(t_a=0,t_b=0)$ by $P(K_S, \bar K^0)$
etc.\ the generalized Bell-CHSH inequality (\ref{dieSgleichung}) gives the following
inequality:
\begin{eqnarray}\label{wigneruchiyama}
P(K_S,\bar K^0)\;\leq\;P(K_S, K^0_1)+P(K^0_1, \bar K^0)\;.
\end{eqnarray}

This is a Wigner-type inequality, originally found by Uchiyama \cite{Uchiyama} by a
set theoretical approach. The interesting point here is its connection to a
physical parameter, the $CP$ violating parameter $\varepsilon$. Calculating the
above probabilities this inequality is turned into an inequality for $\varepsilon$
\begin{eqnarray}\label{uchiyama}
Re\{\varepsilon\}\;\leq\;|\varepsilon|^2.
\end{eqnarray}

The experimental value of $\varepsilon$, measured in experiments, which have nothing to do with
entangled states, has an absolute value of about $10^{-3}$ and a phase of about
$45^\circ$. The quick calculation in the readers head gives - yes - this inequality is violated!
So we have found a way of distinguishing between LRT and QM.

But wait, first we have to see, if this inequality holds also for different times.
For $t=0$ the situation is not really physical, the particles didn't yet fly apart.
So what happens if we let the system evolve in time?

Let us first set all times equal $t_a=t_b=t_c=t$, we get the following inequality
\begin{eqnarray}\label{gleichzeitiguchiyama}
e^{-2 \gamma t}\; P(K_S,\bar K^0)\;\leq\;e^{-2 \gamma t}\; P(K_S, K^0_1)+e^{-2\gamma
t}\; P(K_1^0,\bar K^0)+h(K_S,\bar K^0,K_1^0;t),\nonumber\\
\end{eqnarray}
where $h$ is
\begin{eqnarray}
h(K_S,\bar K^0,K_1^0;t)&=&-P_{K_S,\bar K^0}(N t;N t)+P_{K_S,K^0_1}(N t;N t)
+P_{K_1^0,\bar K^0}(N t;N t)\nonumber\\
& &+P_{K_1^0,K_1^0}(N t;N t)\,.
\end{eqnarray}

This function $h$ is missing, if one does not consider a unitary time
evolution, then the exponential factors $e^{-2 \gamma t}$ can be
divided out of Eq.(\ref{gleichzeitiguchiyama}) and one would wrongly conclude that inequality
(\ref{uchiyama}) holds for all times.

However, it turns out that inequality (\ref{gleichzeitiguchiyama}) is
only for times $t<8\cdot10^{-4}\tau_S$ violated
due to
the fast damping of the probabilities.
Thus for larger times we again can't distinguish between LRT and QM.

But, fortunately, there exist certain cases where the situation is better. We can
avoid a fast increase of the function $h$ by taking the times $t_a=t_c$ and $t_a\leq
t_b$. Then a violation of the Bell-CHSH inequality occurs, which is strongest for $t_a\approx 0$; and in this
case $t_b$ can be chosen up to $t_b\leq 4 \tau_S$, which is really quite large.

The reader agrees for sure that considering Bell inequalities for neutral kaons is a
very strange thing!

\section{Connection of the Bell inequality and decoherence approach}\label{connection}

We have seen in Section \ref{decoherence} that with a simple modification of
the quantum theory we can achieve continuously the factorization of the wave function.
What has this
approach to do with local realistic theories, i.e., Bell inequalities? It is clear
that for $\zeta=0$ - thus quantum theory - Bell inequalities may not be fulfilled, but
for $\zeta=1$, what is clearly a local situation - no interference term exists
between the two amplitudes - Bell inequalities are certainly satisfied.

\begin{center}
\textbf{What can we say for $\zeta$ values between $0$ and $1$?}
\end{center}

Let's consider again inequality (\ref{wigneruchiyama}) and recalculate it with the
modified probabilities, then we will find a bound on $\zeta$. In this way we can
relate the decoherence approach to a local realistic theory.

The result in our `best' basis choice $K_S K_L$ is \cite{H}
\begin{eqnarray}\label{uchiyamazeta}
\frac{Re\{\varepsilon\}-|\varepsilon|^2}{Re\{\varepsilon\}+4 Re^2\{\varepsilon\}+|\varepsilon|^2}
=0.987\;\leq\;\zeta\;.
\end{eqnarray}

So the decoherence parameter $\zeta$ has to be close to one; hence, Furry's hypothesis or
spontaneous factorization has to take place totally. This means in our case that
the created initial state vector (\ref{psikskl})
factorizes in $50\%$ of the cases in a short lived state at the left side and in a
long lived state at the right side or in the other $50\%$ of the cases vice versa.

Intuitively, we would have expected that there exist local realistic theories which
allow at least partially for an interference term, see for instance \cite{Six2,SelleriBook}.
Our result demands for a vanishing
interference term, hence, the locality assumption underlying this inequality forces
the $K_S K_L$ interference term to vanish.

But on the other hand we can compare this result with the experimental
$\bar \zeta^{K_S K_L}=0.13^{+0.16}_{-0.15}$,
Eq.(\ref{fitresultkskl}), where $\zeta=1$ is excluded by many standard
deviations.\\
\\
\textbf{This means that for experimental reasons a local realistic
variable theory equivalent to the $K_S K_L$ basis choice is excluded!}\\
\\
However, the situation changes when using the $K^0 \bar K^0$ basis, then we cannot
discriminate between QM and Furry's hypothesis (for details see Ref.\cite{H}).

\section{Final remark}

The authors hope that they could give a short introduction into
the EPR-Bell-like correlations in particle physics, demonstrating
their similarities to the photon system, but more challenging the
differences (for analogous treatment of the beauty system see for
instance Refs.\cite{BG2,BG3}). We want to emphasize that this
field is a very young one and many experiments are still missing,
but will be done in the near future.

\end{document}